\documentclass[a4paper]{jpconf}
\usepackage{graphicx}
\usepackage{amssymb} 
\usepackage{amsthm} 
\usepackage{lineno}
\usepackage{multirow}
\usepackage{cite}

\newcommand{\DO}{$D^0$ }
\newcommand{\pT}{$p_{T}$ }
\newcommand{\gevc}{$\rm{GeV}/c$ }
\newcommand{\pp}{$p\mbox{-}p$ }
\newcommand{\pPb}{$p$-Pb }
\newcommand{\PbPb}{Pb-Pb }
\newcommand{\pA}{$p$-A }
\newcommand{\raa}{$R_{AA}$ }

\newcommand{\npart}{$N_{part}$ }
\newcommand{\jpsi}{$J/\psi$ }
\newcommand{\sqrtsnn} {$\sqrt{s_{NN}}$ }
\newcommand{\ycm} {$\rm{y_{CM}}$ }

\begin{document}
\title{Heavy flavor Production and Interactions in Relativistic Heavy-Ion Collisions in CMS Experiment}

\author{Wei Xie for the CMS  collaboration}

\address{Department of Physics and Astronomy, PURDUE University, West Lafayette, IN}

\ead{wxie@purdue.edu}

\begin{abstract}
This paper presents the CMS measurements of quarkonia and open heavy flavor production in \pp, \pPb, and \PbPb collisions at \sqrtsnn = 2.76 and 5.02 TeV.  A brief outlook of the near-future CMS heavy flavor physics analyses is provided at the end.
\end{abstract}

\section{Introduction}
        
Measurements of heavy quarkonia and open heavy flavor production may provide further insight in understanding the properties of the strongly coupled Quark-Gluon plasma (sQGP). The quarkonium  production is sensitive to the medium's temperature and the state of deconfinement due to different quarkonium binding energies. Besides the large suppression of production rate caused by mechanisms like the Debye screening~\cite{matsui}, the coalescence process was found to be important for quarkonium  production at RHIC and LHC because of the large number of heavy quarks produced~\cite{recomb}. Open heavy flavor measurements not only serve as a good reference in studying quarkonia, but also provide unique inputs in understanding the parton energy loss in the sQGP medium.
The interaction between heavy quarks and the medium involves different mechanisms ~\cite{rad_eloss,coll_eloss, coll_dissoc, ads_cft} from those of light quarks due to the much larger masses and can be studied through measuring open heavy flavor production.  Furthermore, cold nuclear matter effects (CNM)~\cite{cgc} can also lead to enhancement or suppression of production yield in heavy-ion collisions. To disentangle all these effects, we need to study a wide variety of species of heavy flavor probes in broad \pT ranges in \pp, \pA, and A-A collisions. 

The CMS experiment is a large acceptance, high performance, and multipurpose experiment. 
In CMS, heavy quarkonium  production is measured through dilepton decay channels. The open heavy flavor measurements are done either indirectly through decay daughters or directly through  full reconstruction of $D$, $B$ hadrons, and heavy flavor jets. In this article, Sec.~\ref{heavy_quarkonia_production} describes the CMS heavy quarkonia measurements including charmonia and  bottomonia production  in \PbPb, \pPb, and \pp collisions. Sec.~\ref{open_HF} presents the measurements of open heavy flavor production, including \DO, non-prompt \jpsi and heavy flavor jets in in \PbPb, \pPb, and \pp collisions. At the end of this article, a summary and a brief outlook of the CMS heavy flavor physics analyses in the near future is provided. 

\section{Heavy Quarkonium Measurements}
\label{heavy_quarkonia_production}

\begin{figure}[htbp]
\begin{minipage}[b]{0.33\linewidth}
\begin{center}
\includegraphics[width=\textwidth]{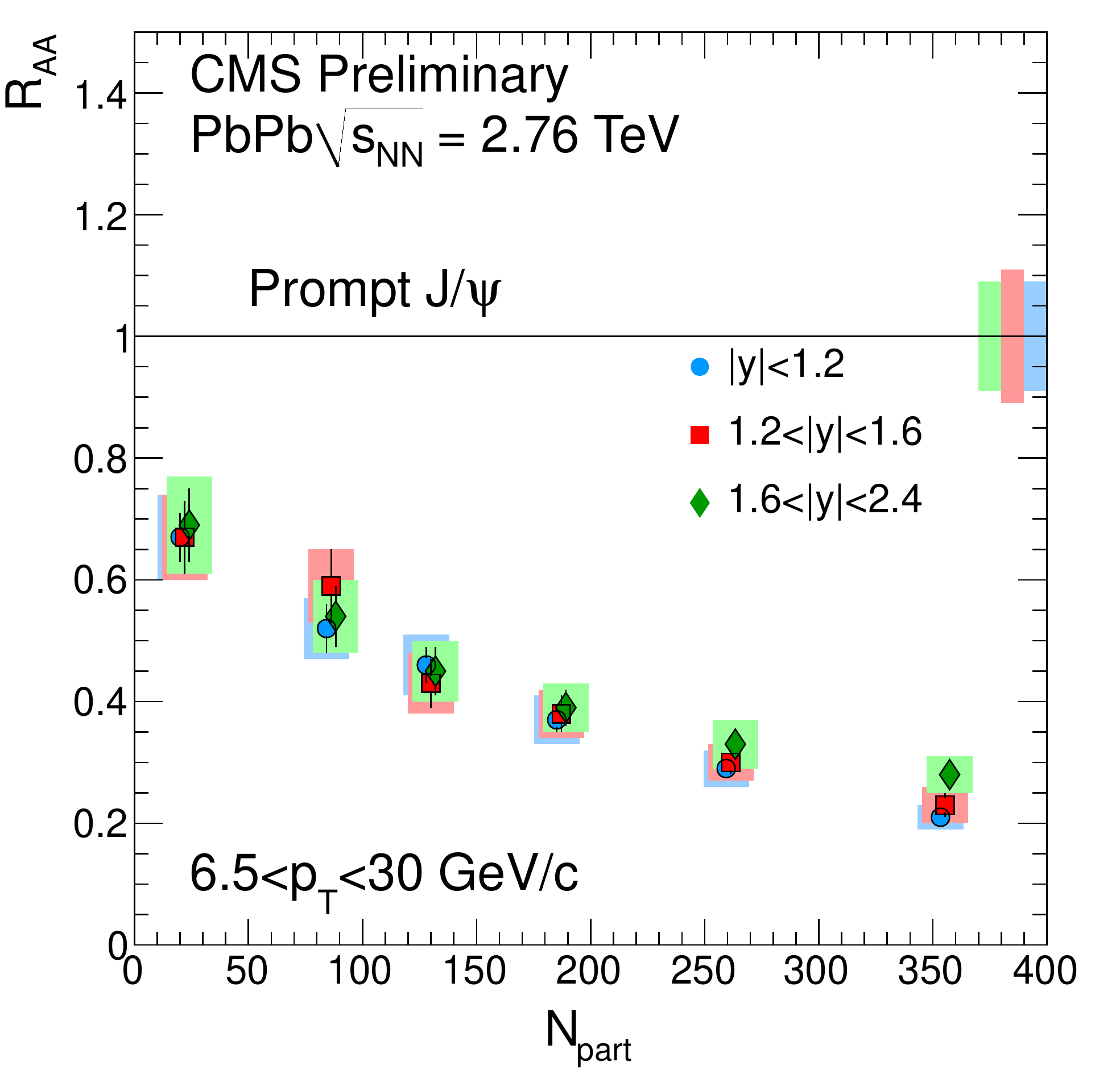}
\end{center}
\end{minipage}
\hspace{-0.1cm}
\begin{minipage}[b]{0.33\linewidth}
\begin{center}
\includegraphics[width=\textwidth]{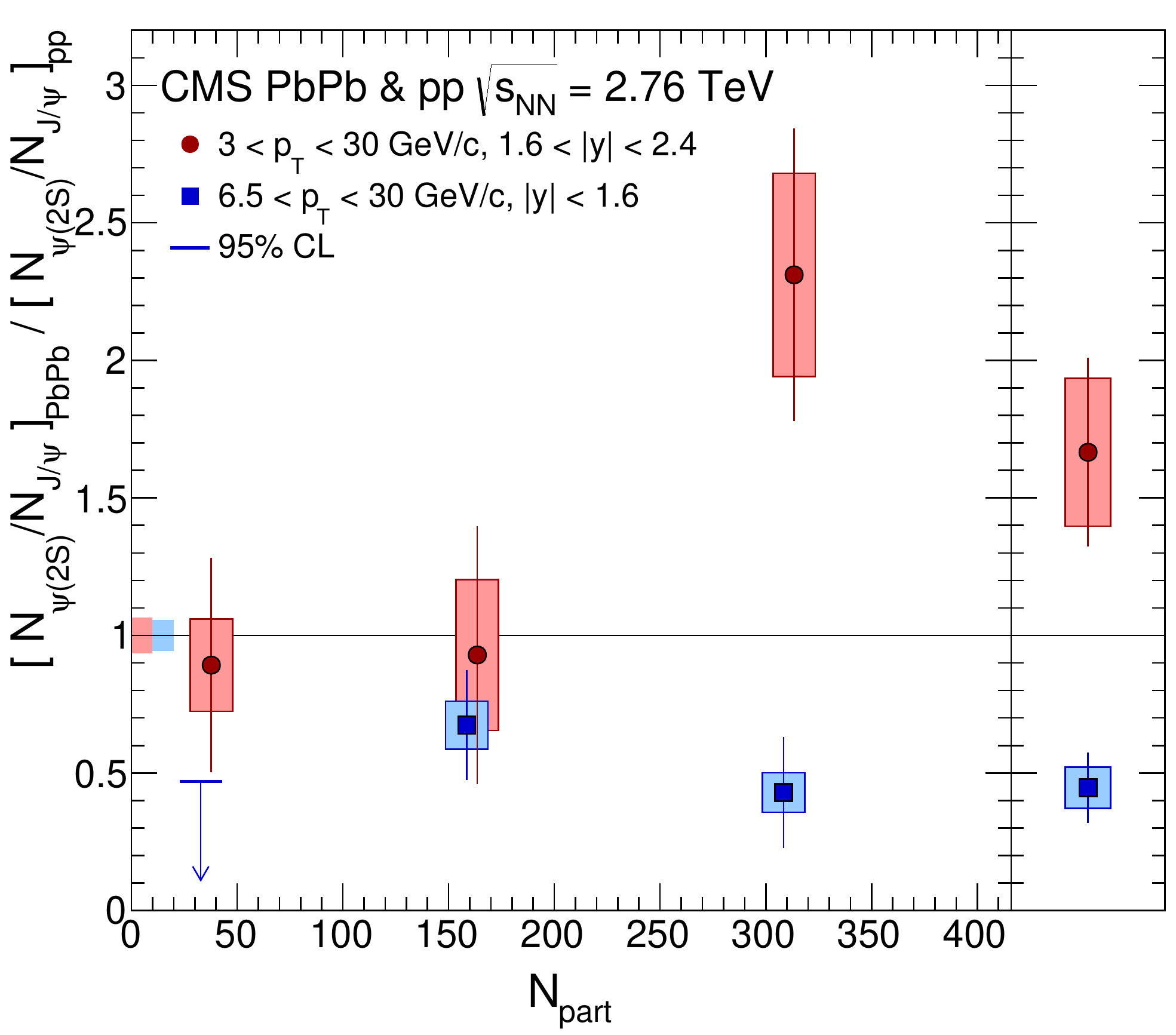}
\end{center}
\end{minipage}
\hspace{-0.1cm}
\begin{minipage}[b]{0.33\linewidth}
\begin{center}
\includegraphics[width=\textwidth]{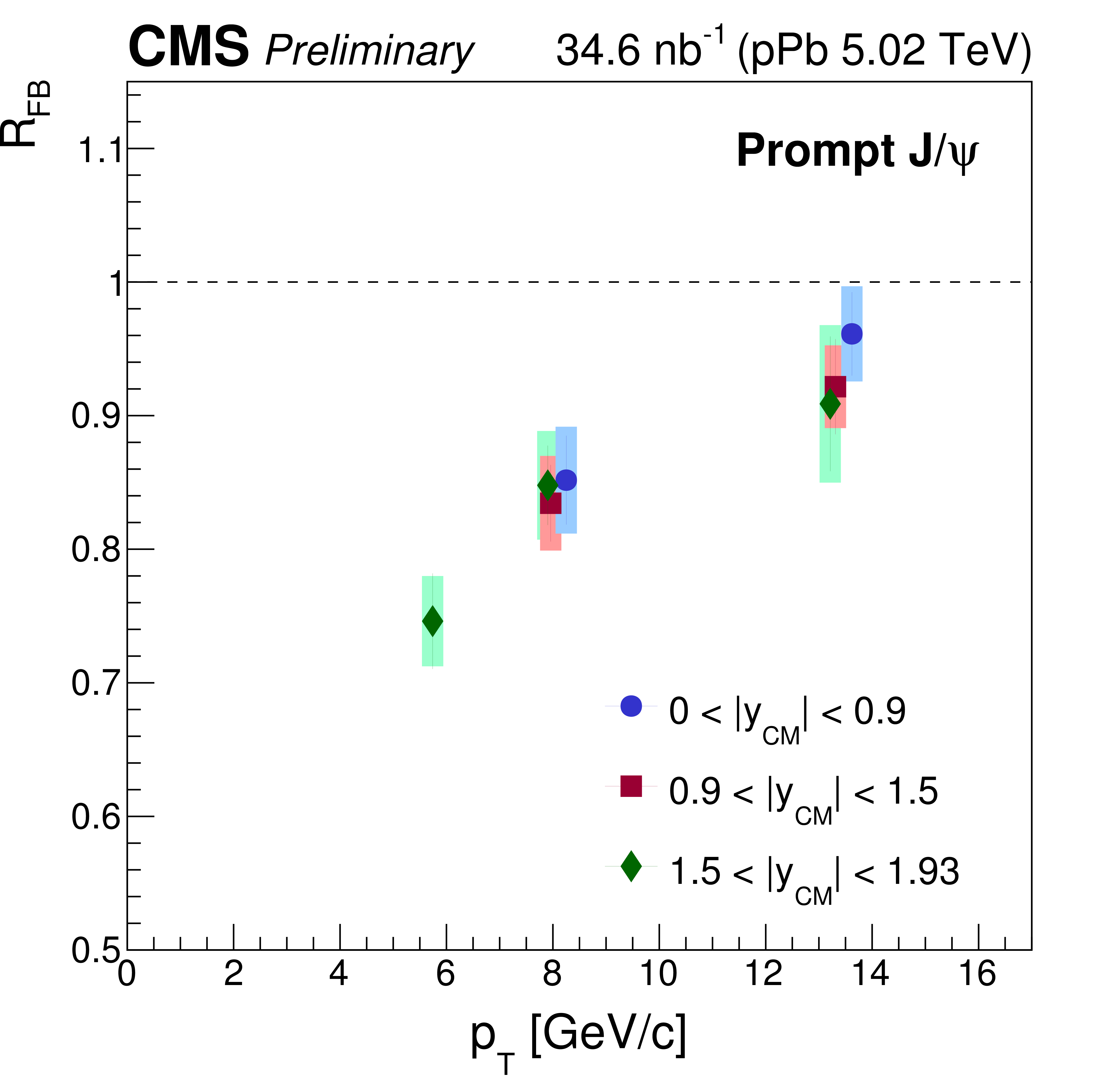}
\end{center}
\end{minipage}
\caption{(Left) Prompt \jpsi \raa as a function of \npart for 6.5 $<$ \pT $<$ 30 \gevc in different rapidity regions. (Middle) Double ratio $[N_{\psi(2S)}/N_{J/\psi}]_{PbPb} / [N_{\psi(2S)}/N_{J/\psi}]_{pp}$ as a function of \npart for 3$<$\pT$<$30 \gevc in 1.6 $< |$y$| <$ 2.4 and for 6.5$<$\pT$<$30 \gevc in $|$y$| <$ 1.6. (Right) Prompt \jpsi $R_{FB}$ as a function of \pT in different \ycm regions.}
\label{fig:jpsiPsi}
\end{figure}

The left panel of Figure~\ref{fig:jpsiPsi} shows the prompt \jpsi \raa as a function of \npart at high \pT (6.5$<$\pT$<$ 30 \gevc) measured within three rapidity regions in 2.76 TeV \PbPb collisions~\cite{HIN-12-014}. Compared to the low \pT measurements at the same energy~\cite{ALICE_jpsi}, high \pT \jpsi is more suppressed. This \pT dependence is opposite to what is observed at RHIC  in Au-Au collisions at \sqrtsnn = 200 GeV~\cite{RHIC_jpsi} and may suggest interplay among suppression, regeneration from coalescence in sQGP medium and  CNM effects in different collision energies. The middle panel shows the double ratio  
$[N_{\psi(2S)}/N_{J/\psi}]_{PbPb} / [N_{\psi(2S)}/N_{J/\psi}]_{pp}$ 
as a function \npart for two different \pT regions. The results indicate $\psi(2S)$ is less suppressed than \jpsi for low \pT (3$<$\pT$<$30 \gevc) and more suppressed for higher \pT region (6.5$<$\pT$<$30 \gevc), which is possibly caused by the coalescence production at different time scale during the medium evolution due to different binding energies ~\cite{seq_reg}. The right panel shows the prompt \jpsi 
$R_{FB}= \frac{d^2\sigma(\rm{y}_{CM} > 0)/dp_Td\rm{y}_{CM}}{d^2\sigma(\rm{y}_{CM} < 0)/dp_Td\rm{y}_{CM}}$ as a function of \pT in 5.02 TeV \pPb collisions, where positive \ycm is defined as the proton-going direction. The prompt \jpsi production in \ycm $>$ 0 region is found to be less than that in \ycm $<$ 0 region, presumbly due to CNM effects. The $R_{pA}$ of prompt \jpsi is needed and is expected to be ready soon to tell how CNM effects play a role in \raa measurements in \PbPb collisions. CMS also measured the coherent \jpsi photoproduction in ultra-peripheral \PbPb collisions at \sqrtsnn = 2.76 TeV~\cite{UPC}. The results are consistent with the leading twist approximation which includes nuclear gluon shadowing.     

\begin{figure}[htbp]
\begin{minipage}[b]{0.33\linewidth}
\begin{center}
\includegraphics[width=\textwidth]{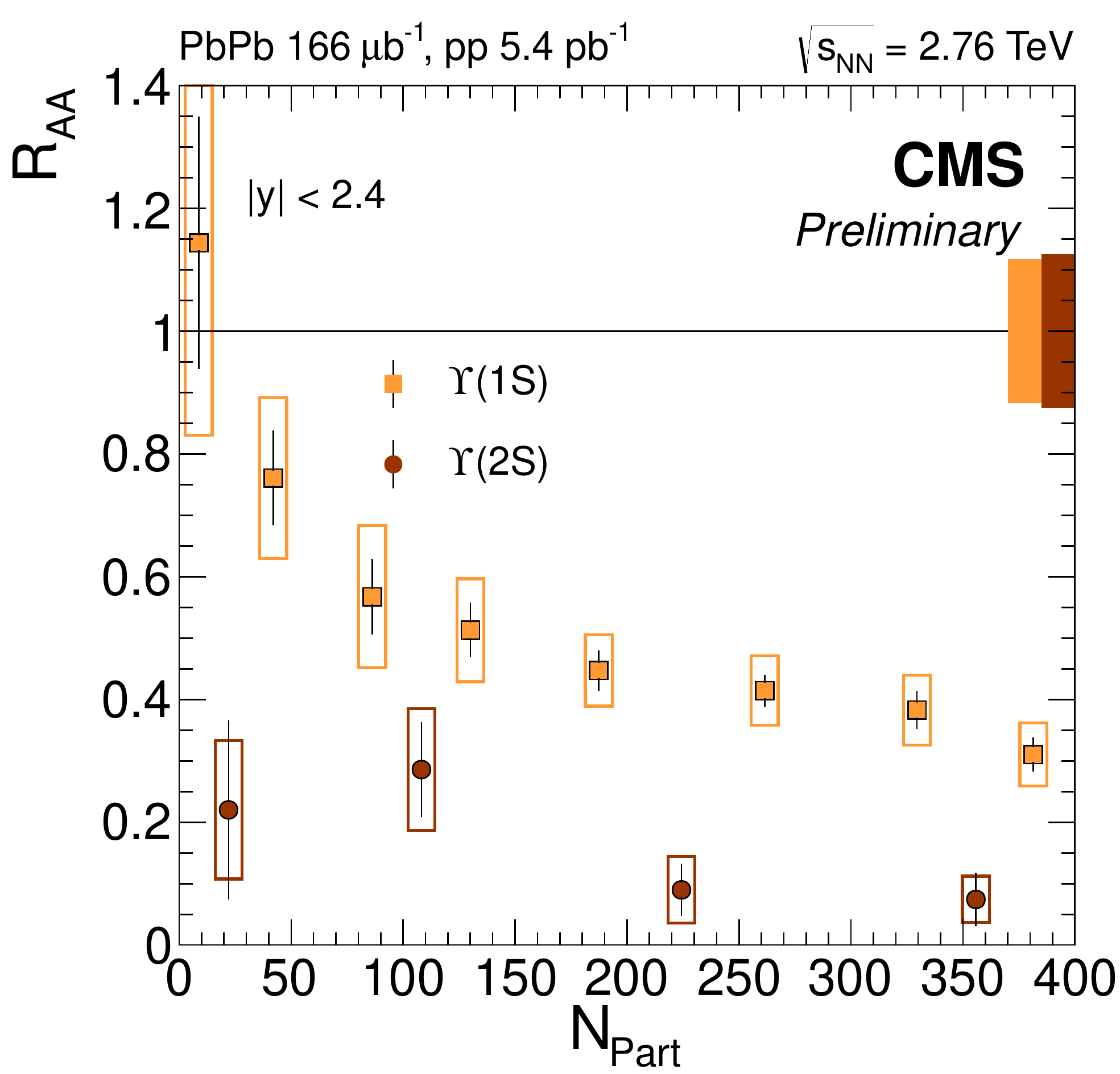}
\end{center}
\end{minipage}
\hspace{-0.1cm}
\begin{minipage}[b]{0.33\linewidth}
\begin{center}
\includegraphics[width=\textwidth]{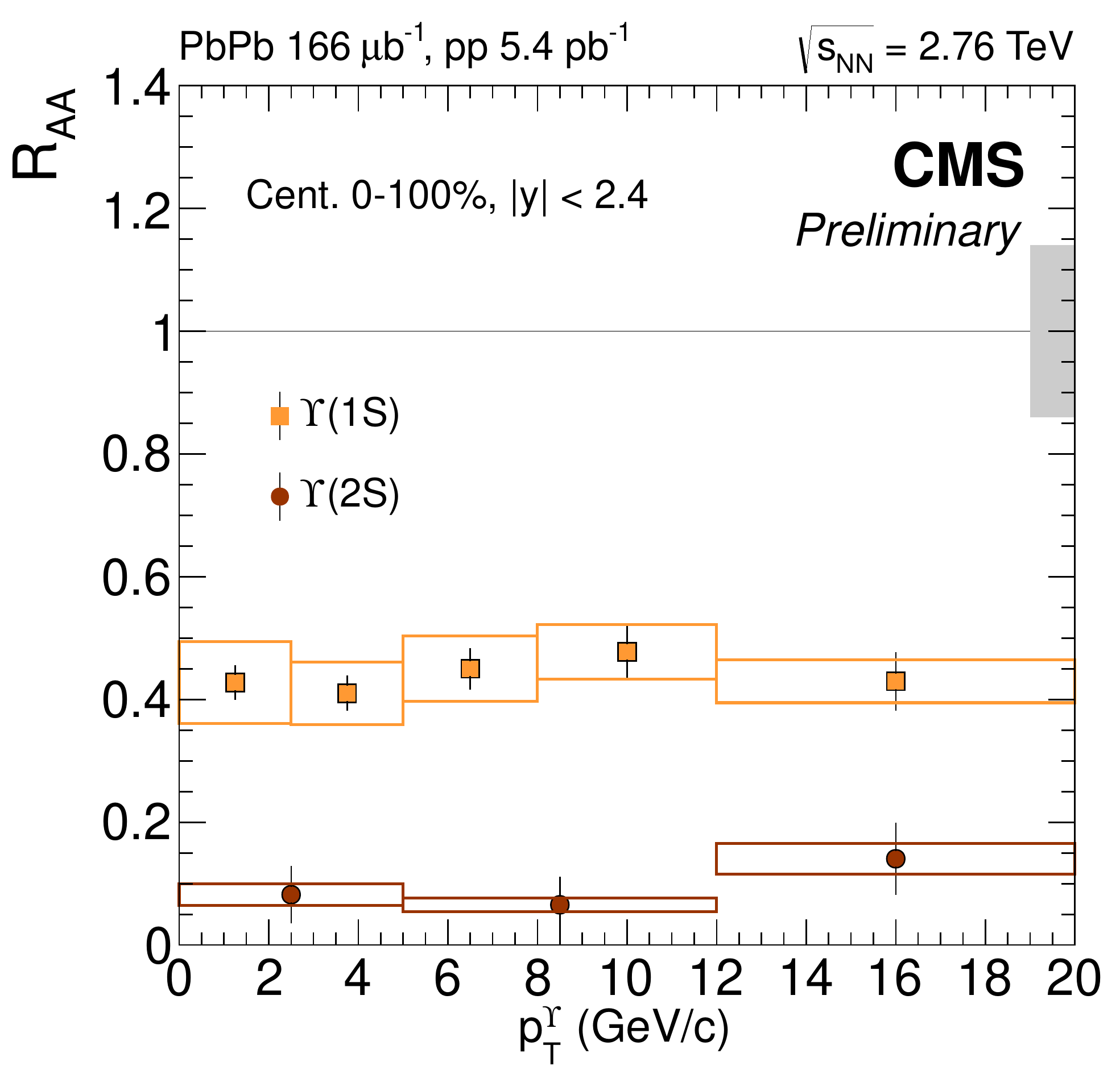}
\end{center}
\end{minipage}
\hspace{-0.1cm}
\begin{minipage}[b]{0.33\linewidth}
\begin{center}
\includegraphics[width=\textwidth]{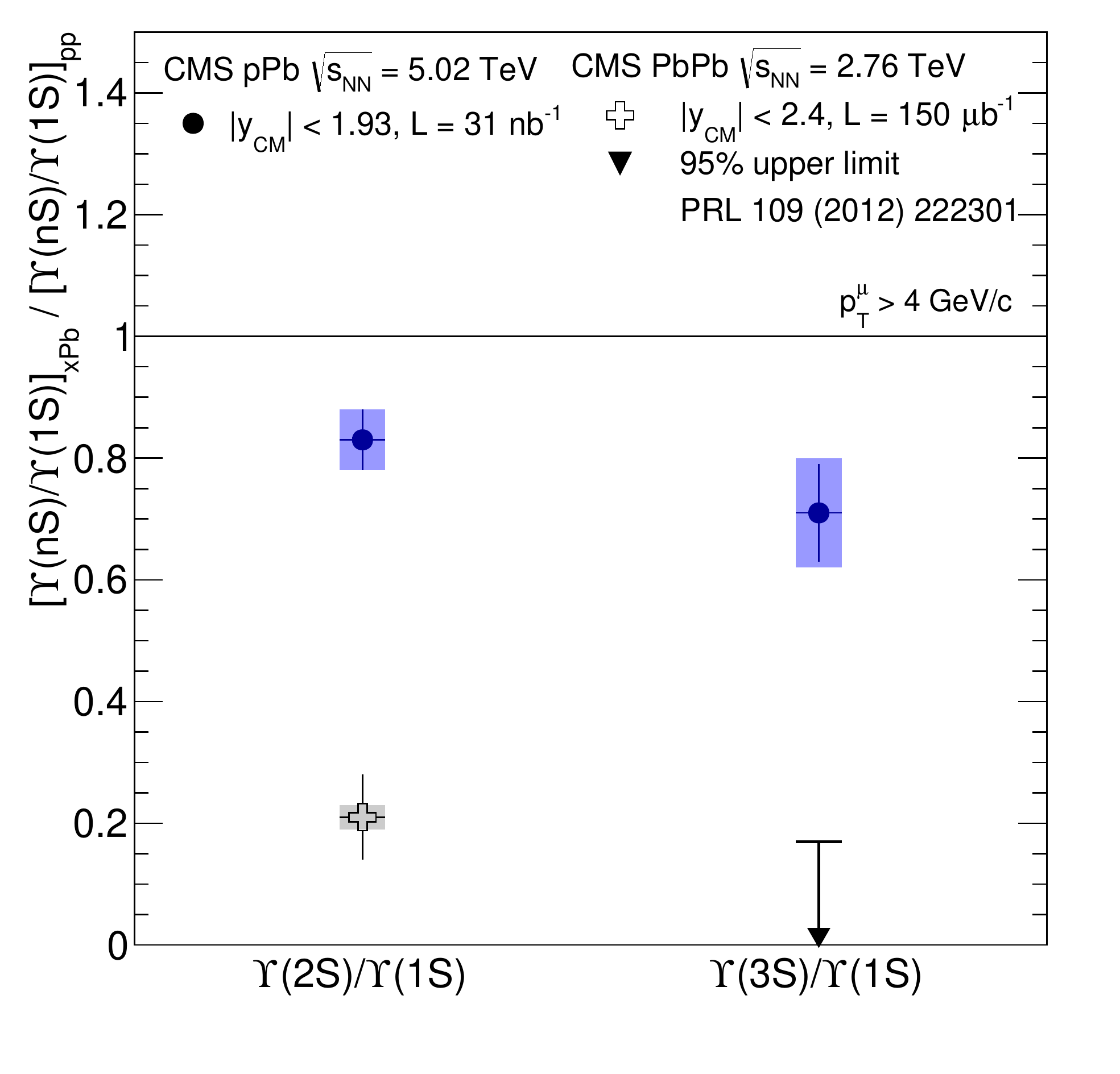}
\end{center}
\end{minipage}
\caption{(Left) \raa as a function of \npart for $\Upsilon(1S)$ and $\Upsilon(2S)$ in \PbPb collisions. (Middle) \raa as a function of \pT for $\Upsilon (1S)$ and $\Upsilon(2S)$ in \PbPb collisions. (Right) Double ratio $[\Upsilon (nS)/\Upsilon (1S)]_{PbPb}/[\Upsilon (nS)/\Upsilon (1S)]_{pp}$ for $\Upsilon (2S)$ and $\Upsilon (3S)$ in \pPb and \PbPb collisions.}
\label{fig:upsilonNs}
\end{figure}

In Figure~\ref{fig:upsilonNs}, the left and middle panels show the $\Upsilon(nS)$ \raa as a function of \npart and \pT in 2.76 TeV \PbPb collisions, respectively~\cite{HIN-15-001}. The \raa of $\Upsilon(nS)$ are found to be ordered with the binding energy of each state for all centralities and for all \pT regions. The suppression  in \PbPb collisions is mostly from the sQGP medium effect since the double ratio $\frac{\Upsilon (nS)/\Upsilon (1S)]_{PbPb}}{\Upsilon (nS)/\Upsilon (1S)]_{pp}}$ observed in \pPb collisions is much higher than that in \PbPb collisions as shown in the right panel. Contrary to what is observed from the $\psi(2S)$ and \jpsi double ratio measurement at low \pT, $\Upsilon(2S)$ is more suppressed than $\Upsilon(1S)$ for all \pT. These different trends may come from the different roles the regeneration contribution plays in $\Upsilon$ and \jpsi production due to their large difference in mass.   

\begin{figure}[htbp]
\begin{minipage}[b]{0.33\linewidth}
\begin{center}
\includegraphics[width=\textwidth]{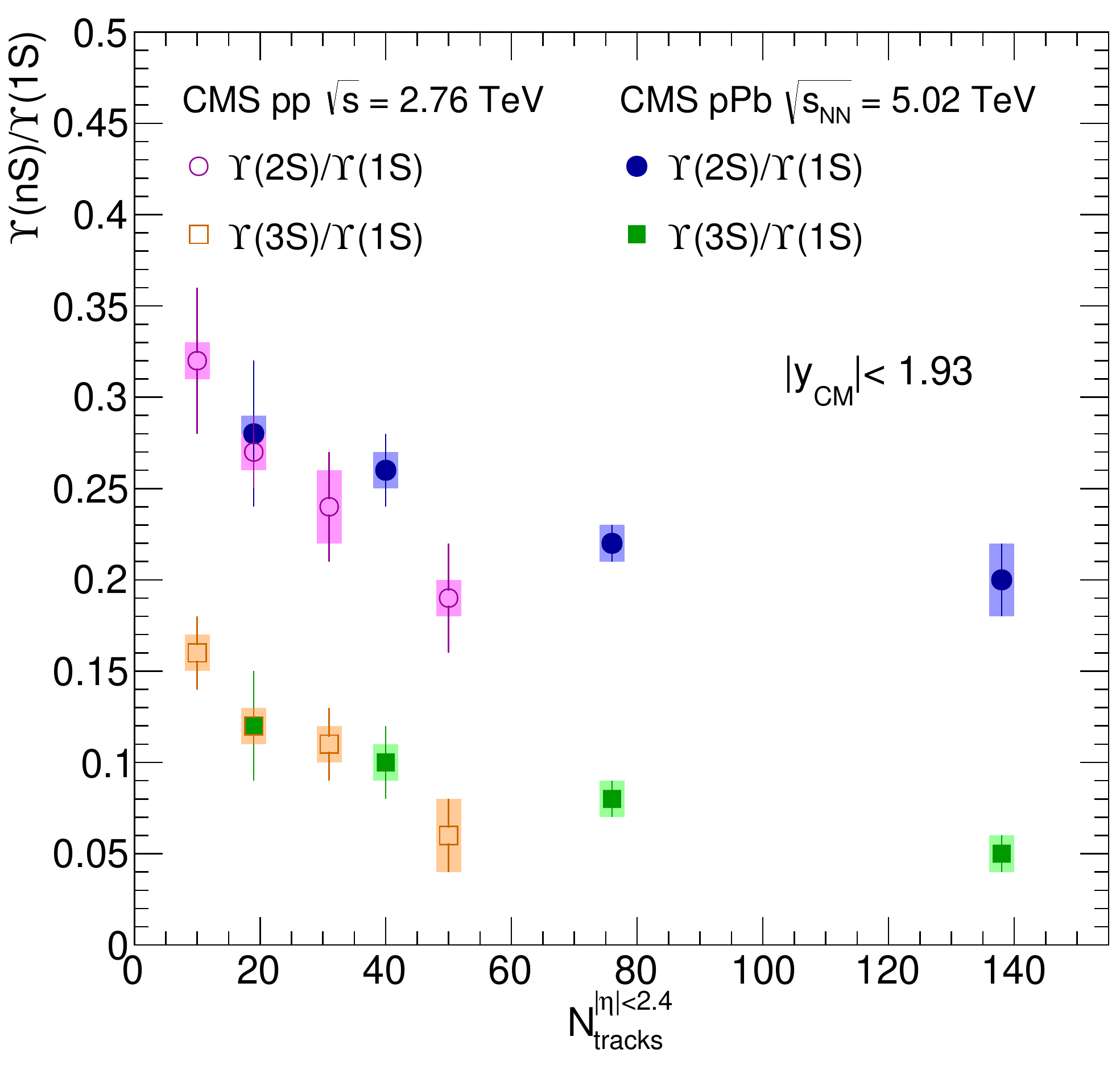}
\end{center}
\end{minipage}
\hspace{-0.1cm}
\begin{minipage}[b]{0.33\linewidth}
\begin{center}
\includegraphics[width=\textwidth]{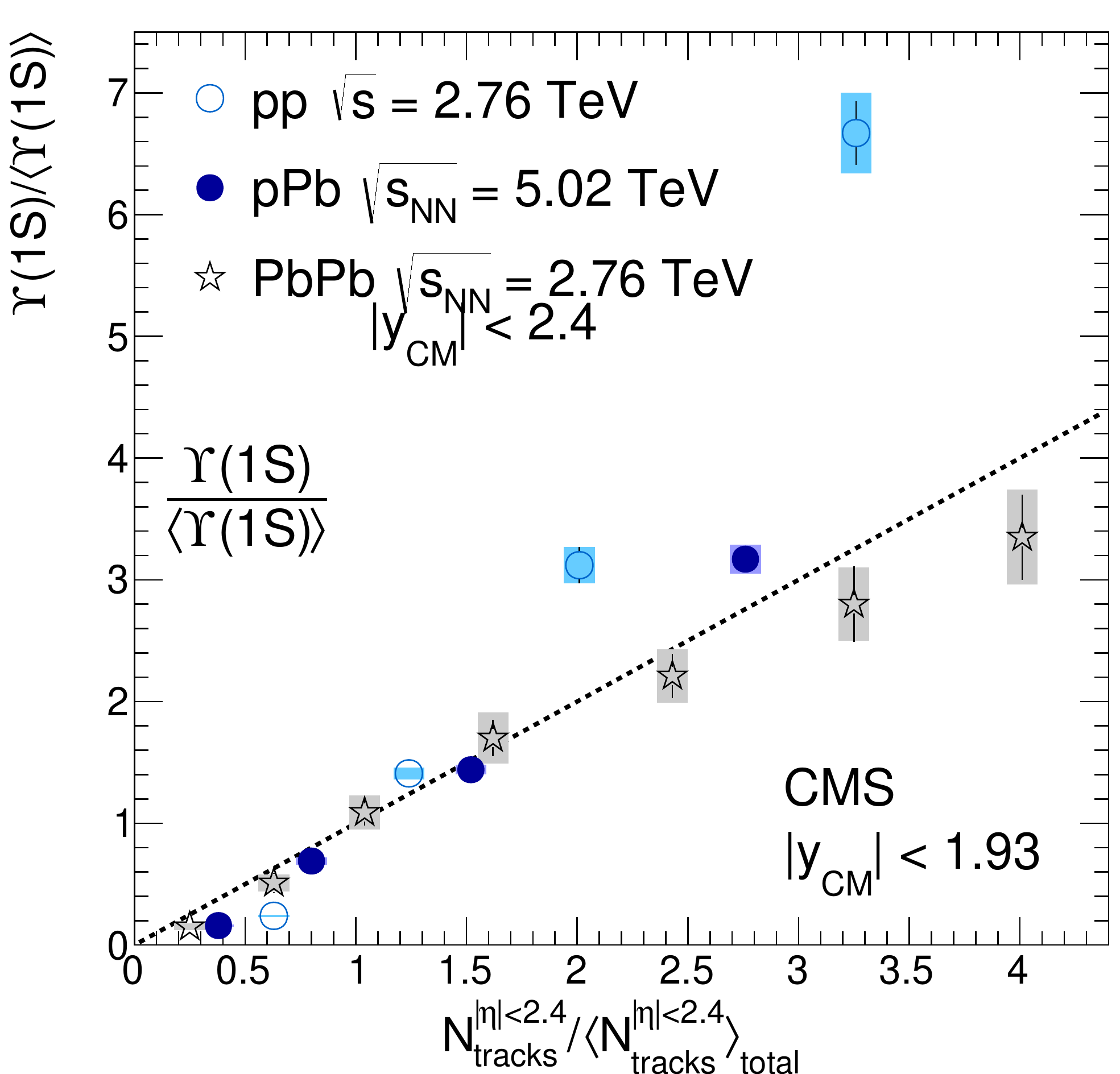}
\end{center}
\end{minipage}
\hspace{-0.1cm}
\begin{minipage}[b]{0.33\linewidth}
\begin{center}
\includegraphics[width=\textwidth]{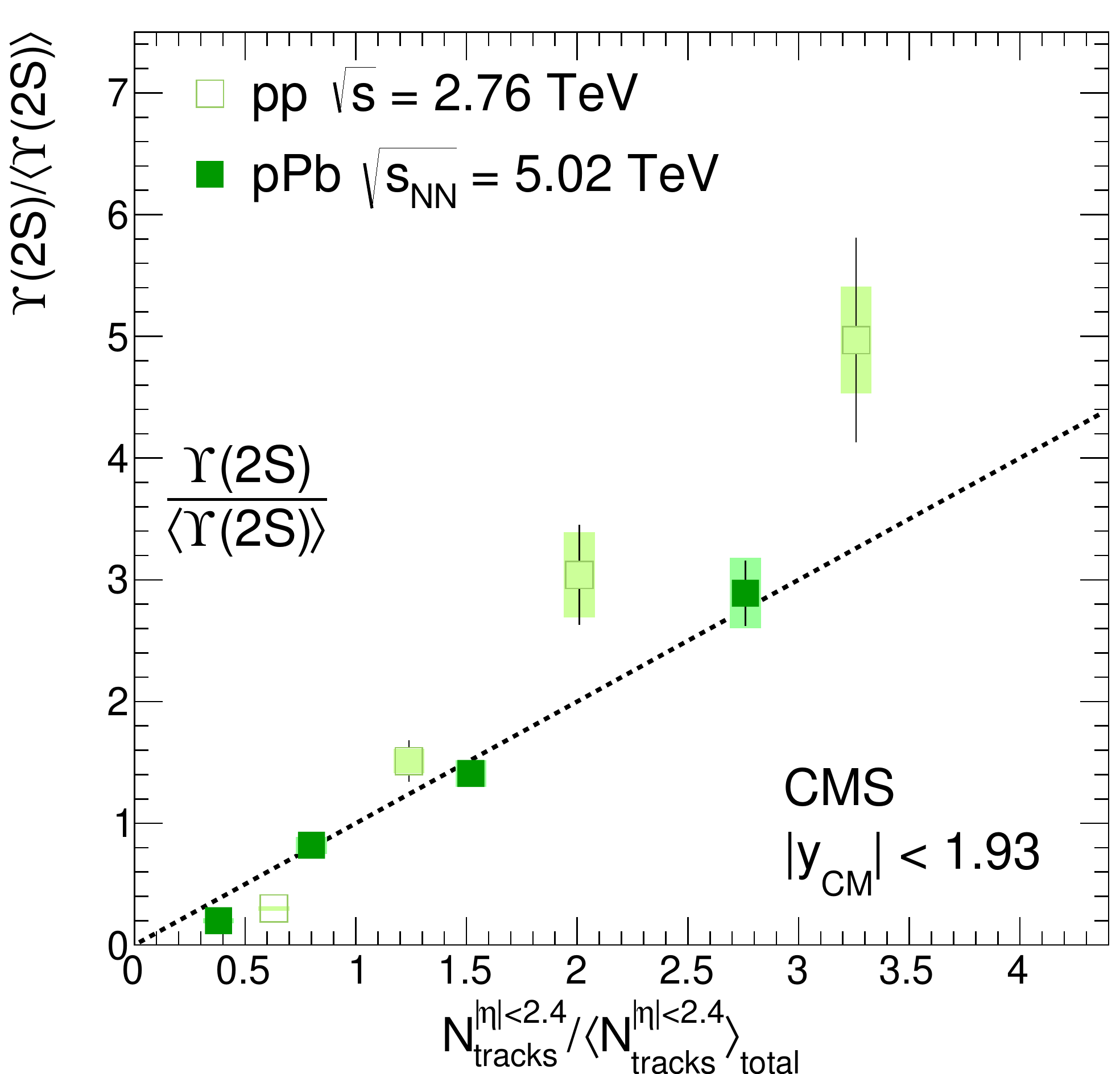}
\end{center}
\end{minipage}
\caption{(Left) Ratio $\Upsilon(nS)/\Upsilon(1S)$ as a function of $N^{|\eta|<2.4}_{tracks}$ in 2.76 TeV \pp and 5.02 TeV \pPb collisions for $\Upsilon(2S)$ and $\Upsilon(3S)$ in $|\rm{y}_{CM}|<$1.93. (Middle) Normalized $\Upsilon(nS)$ cross section as a function of the normalized charged track multiplicity for $\Upsilon(1S)$ in 2.76 TeV \pp, 5.02 TeV \pPb, and 2.76 TeV \PbPb collisions for $|\rm{y}_{CM}|<1.93$. (Right) Same as middle panel but for $\Upsilon(2S)$.}
\label{fig:upsilon_pPb}
\end{figure}

The left panel of Figure~\ref{fig:upsilon_pPb} shows the ratio of $\Upsilon(nS)$ cross section to $\Upsilon(1S)$ cross section, i.e. $\Upsilon(nS)/\Upsilon(1S)$, in \pp and \pPb collisions  as a function of the charged tracks multiplicity in $|\eta| <$ 2.4, where the $\Upsilon$ signals are selected in $|$\ycm$| <$ 1.93~\cite{JHEP04_103}. The $\Upsilon(nS)/\Upsilon(1S)$ is found to decrease toward higher multiplicity. Similar trend is observed at RHIC for $\psi(2S)$ and \jpsi production in $d$-Au collisions~\cite{phnx_jpsi_dAu}. The more suppression of excited states with weaker binding energy indicate final state effects in quarkonia production in high multiplicity \pp and \pPb or $d$-Au collisions. 

\begin{figure}[htbp]
\begin{center}
\includegraphics[width=\textwidth]{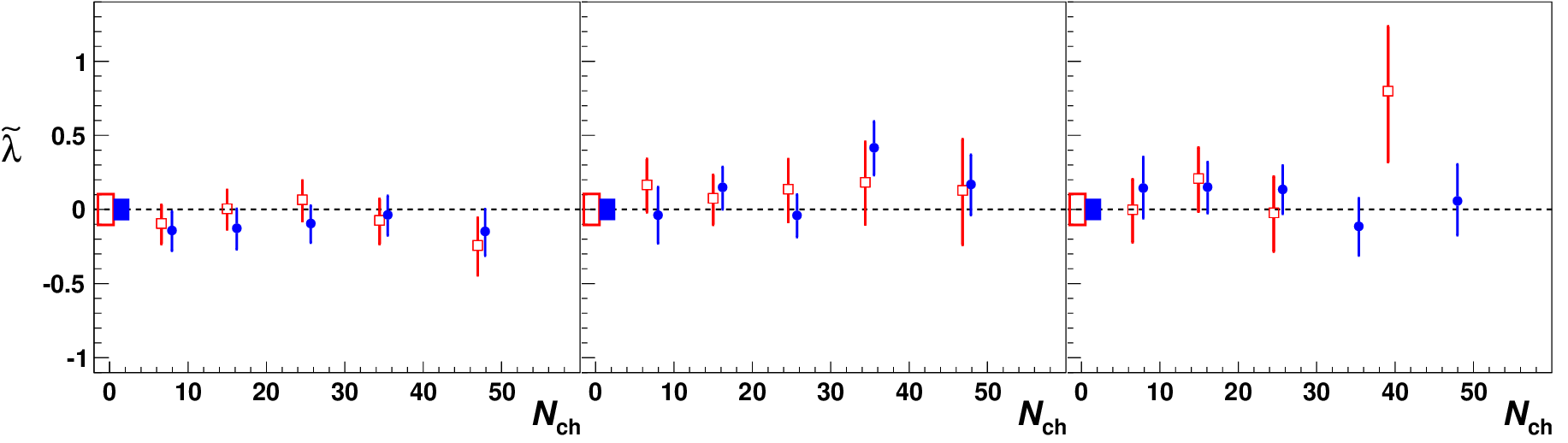}
\end{center}
\caption{The $\tilde{\lambda}$ parameter for the $\Upsilon(1S)$ (left), $\Upsilon(2S)$ (middle), and $\Upsilon(3S)$ (right) states in the center-of-mass helicity frame as a function of $N_{ch}$ for \pT = 10-15 \gevc (open squares) and 15-35 \gevc (closed circles). The boxes at the zero horizontal line represent the global uncertainties. }
\label{fig:upsilon_pol}
\end{figure}

Since a $Q\bar{Q}$ interact with the surrounding QCD medium before forming a quarkonium state, it is essential to understand how this interaction may affect quarkonia formation mechanism and thus the production rate in high multiplicity \pA collisions. A measurement of quarkonia polarization as a function of charged particle multiplicity can provide the needed clues and CMS took the first step toward that direction. Figure~\ref{fig:upsilon_pol} shows measrements of the frame-invariant parameter $\tilde{\lambda}$, in 7 TeV \pp collisions, as a function of charged particle multiplicity in the center-of-mass helicity frame for the three $\Upsilon$ states in 10 $<$ \pT $<$ 15 \gevc and 15 $<$ \pT $<$ 35 \gevc~\cite{Upsilon_polarization}. No significant variation of $\Upsilon$ polarization as a function of charged particle multiplicity  is observed for the two high \pT bins.

\section{Open Heavy Flavor Measurements}
\label{open_HF}

\begin{figure}[htbp]
\begin{minipage}[b]{0.50\linewidth}
\begin{center}
\includegraphics[width=\textwidth]{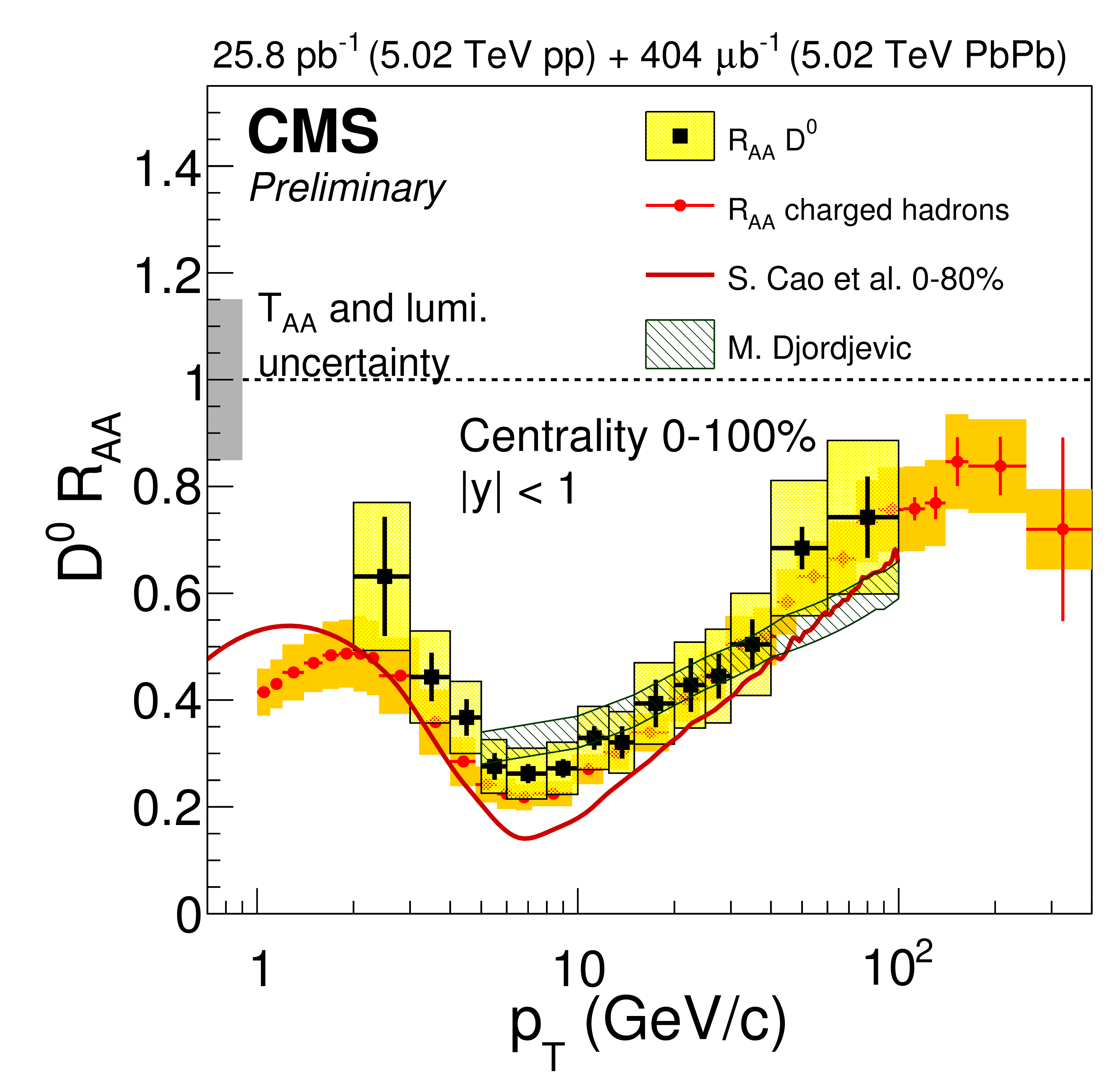}
\end{center}
\end{minipage}
\hspace{-0.1cm}
\begin{minipage}[b]{0.50\linewidth}
\begin{center}
\includegraphics[width=\textwidth]{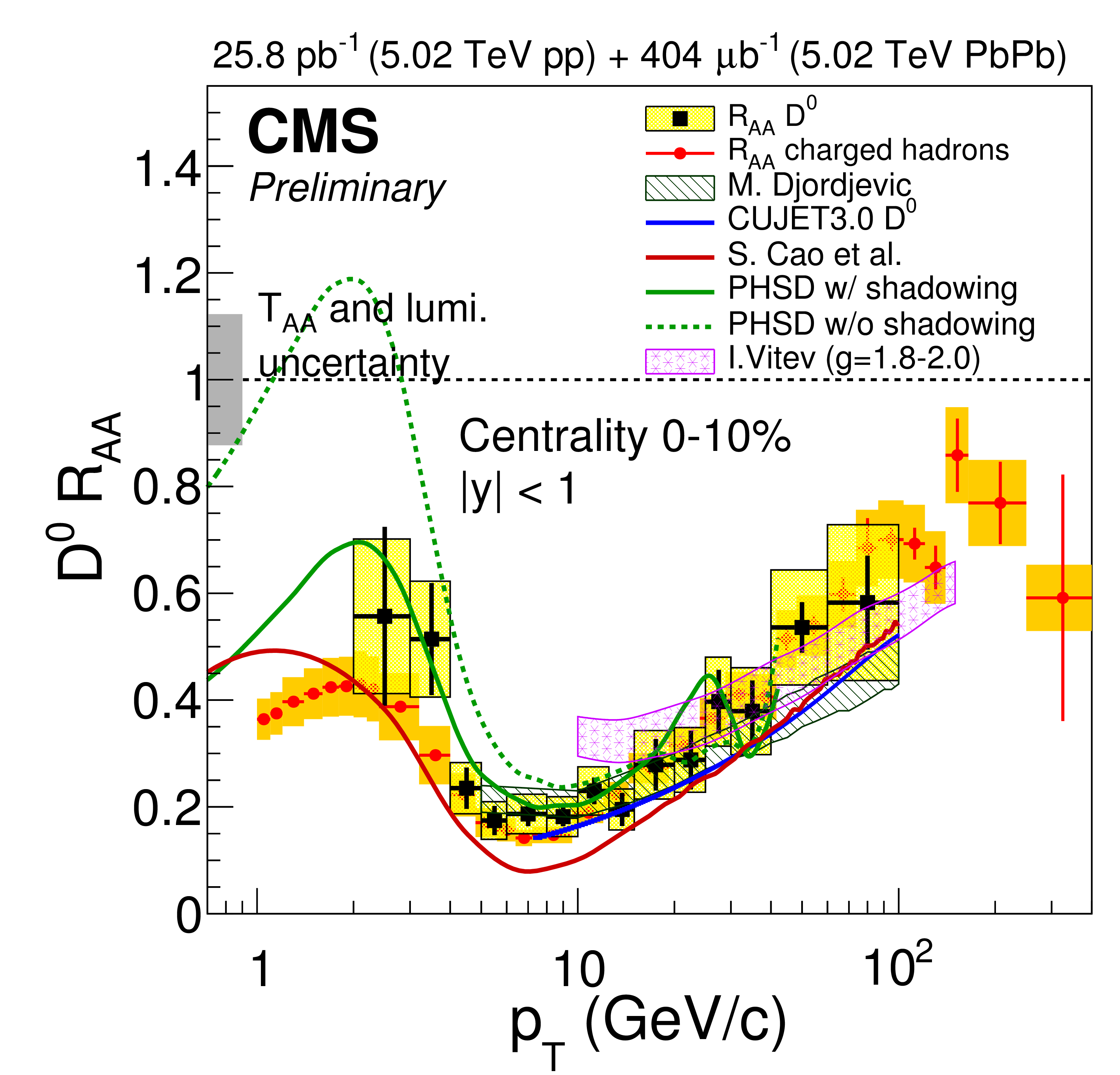}
\end{center}
\end{minipage}
\caption{$D^0$ \raa as a function of \pT in 0-100\% (left) and 0-10\% (right) centrality selection of \PbPb collision at \sqrtsnn = 5.02 TeV. The charged particle \raa are shown for comparison. The bands and curves are model predictions.}
\label{fig:D0_RAA}
\end{figure}

Figure~\ref{fig:D0_RAA} shows the prompt \DO \raa as a function of \pT within $|$y$|<$1.0 in 0-100\% and 0-10\% centrality selection of \PbPb collisions at \sqrtsnn = 5.02 TeV~\cite{HIN-16-001}. One significant improvement over the previous measurements at 2.76 TeV~\cite{HIN-15-005} is that the reference of the new \raa results is from the \pp measurement at the same collisions energy. The prompt \DO \raa is found to decrease with \pT for \pT $<$ 10 \gevc and increase toward higher \pT, which can be qualitatively described by various model calculations~\cite{D0_models}. The charged particle \raa is found to be consistent with \DO \raa within uncertainties for \pT = 2-100 \gevc despite the much larger charm quark mass.  

\begin{figure}[htbp]
\begin{minipage}[b]{0.33\linewidth}
\begin{center}
\includegraphics[width=\textwidth]{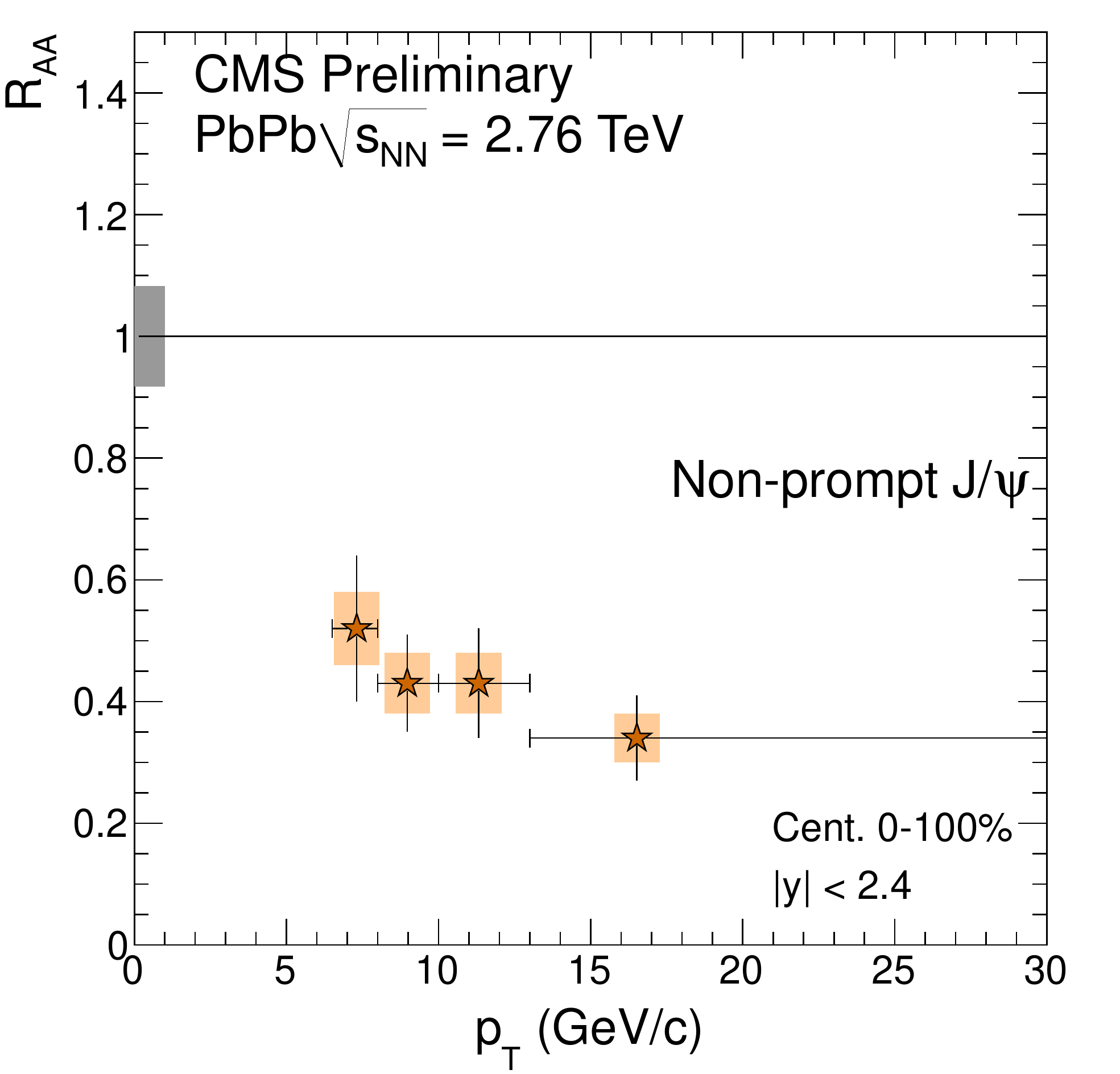}
\end{center}
\end{minipage}
\hspace{-0.1cm}
\begin{minipage}[b]{0.33\linewidth}
\begin{center}
\includegraphics[width=\textwidth]{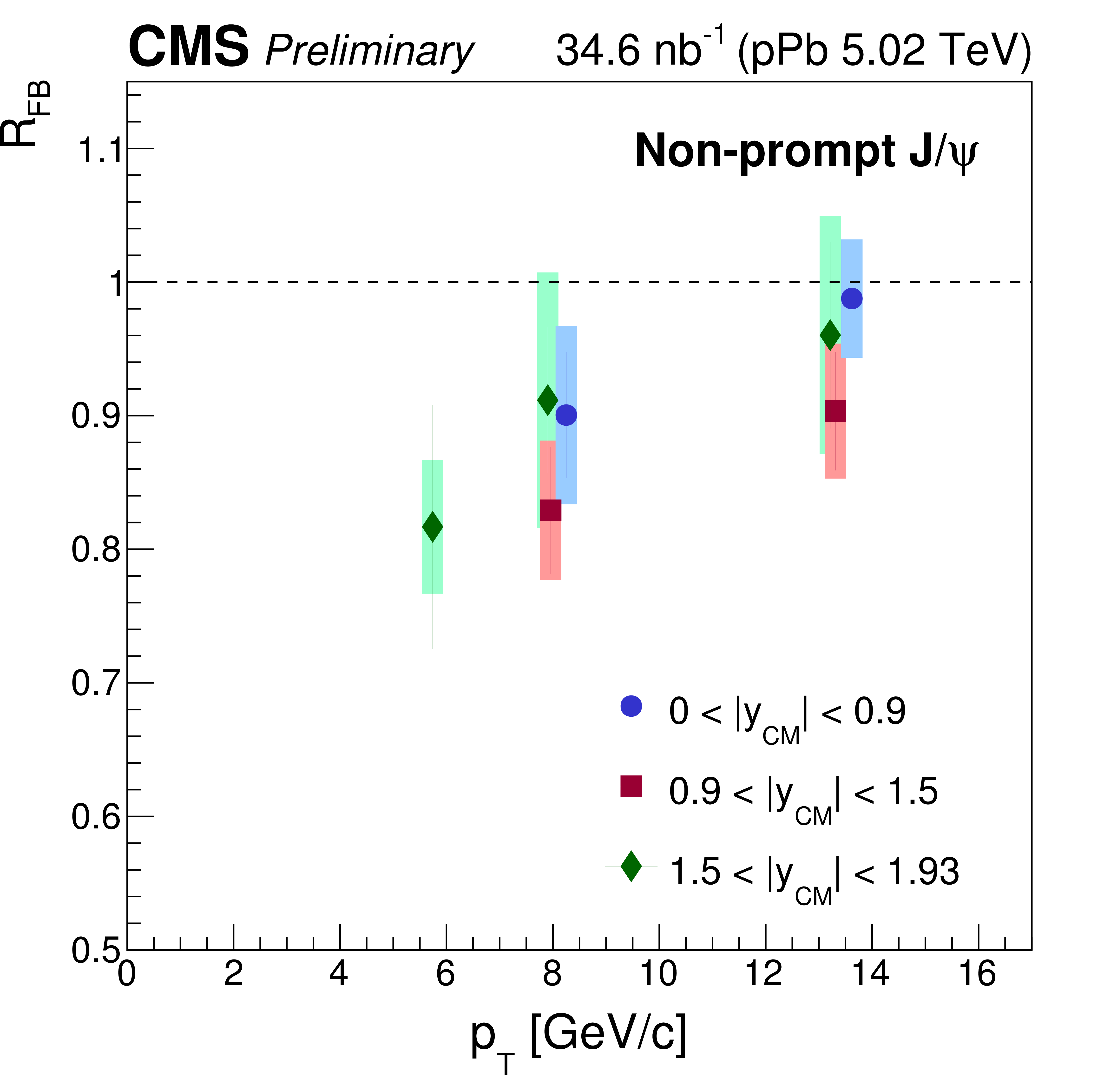}
\end{center}
\end{minipage}
\hspace{-0.1cm}
\begin{minipage}[b]{0.33\linewidth}
\begin{center}
\includegraphics[width=\textwidth]{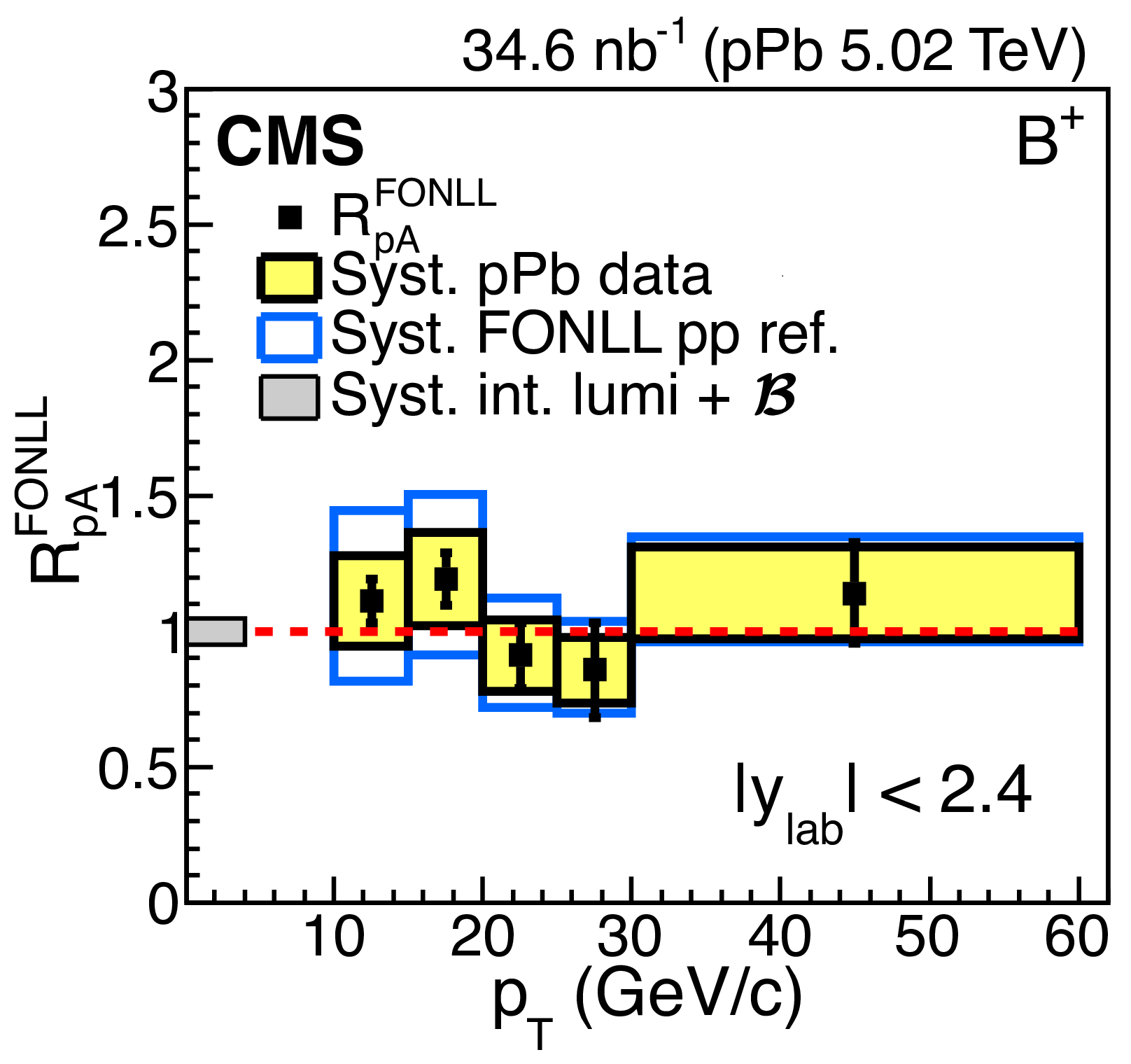}
\end{center}
\end{minipage}
\caption{(Left) Non-prompt \jpsi \raa as a function of \pT in 0-100\% centrality selection of \PbPb collisions at \sqrtsnn = 2.76 TeV. (Middle) Non-prompt \jpsi $R_{FB}$ as a function of \pT in three different \ycm  regions in 5.02 TeV \pPb collisions. (Right) $B^+$ meson $R^{FONLL}_{pA}$ in $|\rm{y}_{lab}| <$ 2.4 as a function of \pT in 5.02 TeV \pPb collisions.}
\label{fig:np_jpsi}
\end{figure}

To further study the mass dependence of energy loss in sQGP medium,  the non-prompt \jpsi from $B$-hadron decays is measured. The left panel of Figure~\ref{fig:np_jpsi} shows non-prompt \jpsi \raa as a function of \pT in 2.76 TeV \PbPb collisions in 0-100\% centrality region for 6.5 \gevc $<$ \pT $<$ 30 \gevc~\cite{HIN-12-014}. For non-prompt \jpsi with \pT $>$ 6.5 \gevc, the parent $B$-hadron \pT is roughly above 8.0 \gevc. The \DO \raa measurement with \pT $>$ 8.0 \gevc can thus be compared to the non-prompt \jpsi results. With large uncertainties, the comparison indicates larger non-prompt \jpsi \raa than that of \DO at \pT $\sim$ 10 \gevc.  Further information can be provided when the non-prompt \jpsi at 5.02 TeV becomes available to compare with the more precise \DO result at 5.02 TeV. The right panel of Figure~\ref{fig:np_jpsi} shows the $B^+$ $R^{FONLL}_{pA}$ as a function of \pT in 5.02 TeV \pPb collisions, where $B^+$ signal is obtained via full reconstruction and the reference is from FONLL calculation~\cite{B_RpPb}. No significant CNM effects are observed within uncertainties in $B$ hadron production. However, the more differential $R_{FB}$ measurements from non-prompt \jpsi in \pPb shows that the production in the \ycm $>$ 0 region is less than that in the \ycm $<$ 0 region at \pT $\sim$ 6 \gevc, as observed in the prompt \jpsi measurements.  
  
\begin{figure}[htbp]
\begin{minipage}[b]{0.33\linewidth}
\begin{center}
\includegraphics[width=\textwidth]{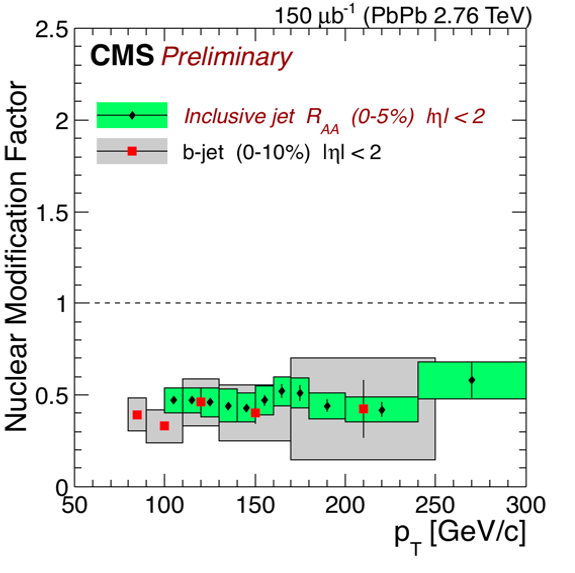}
\end{center}
\end{minipage}
\hspace{-0.1cm}
\begin{minipage}[b]{0.33\linewidth}
\begin{center}
\includegraphics[width=\textwidth]{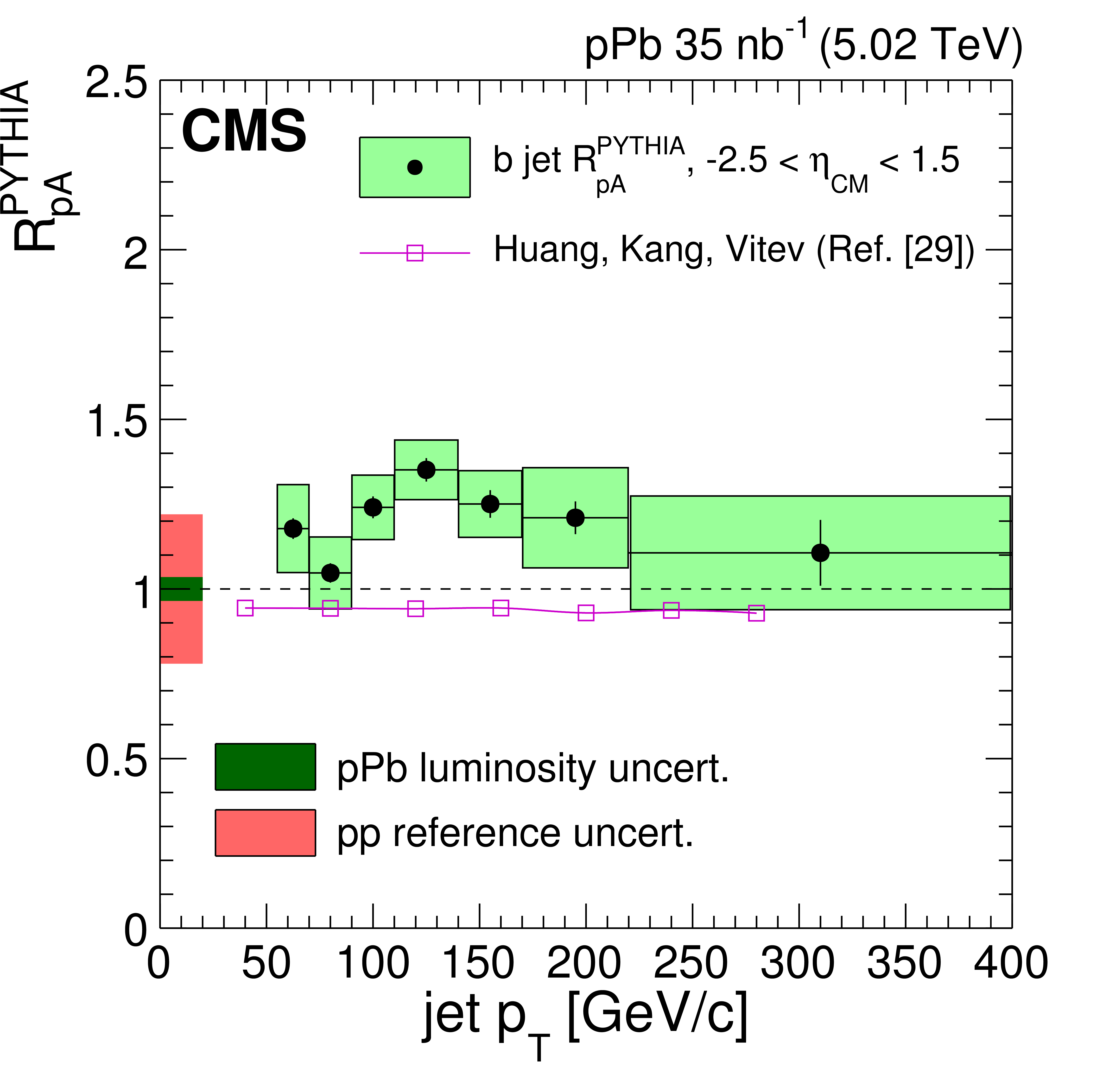}
\end{center}
\end{minipage}
\hspace{-0.1cm}
\begin{minipage}[b]{0.33\linewidth}
\begin{center}
\includegraphics[width=\textwidth]{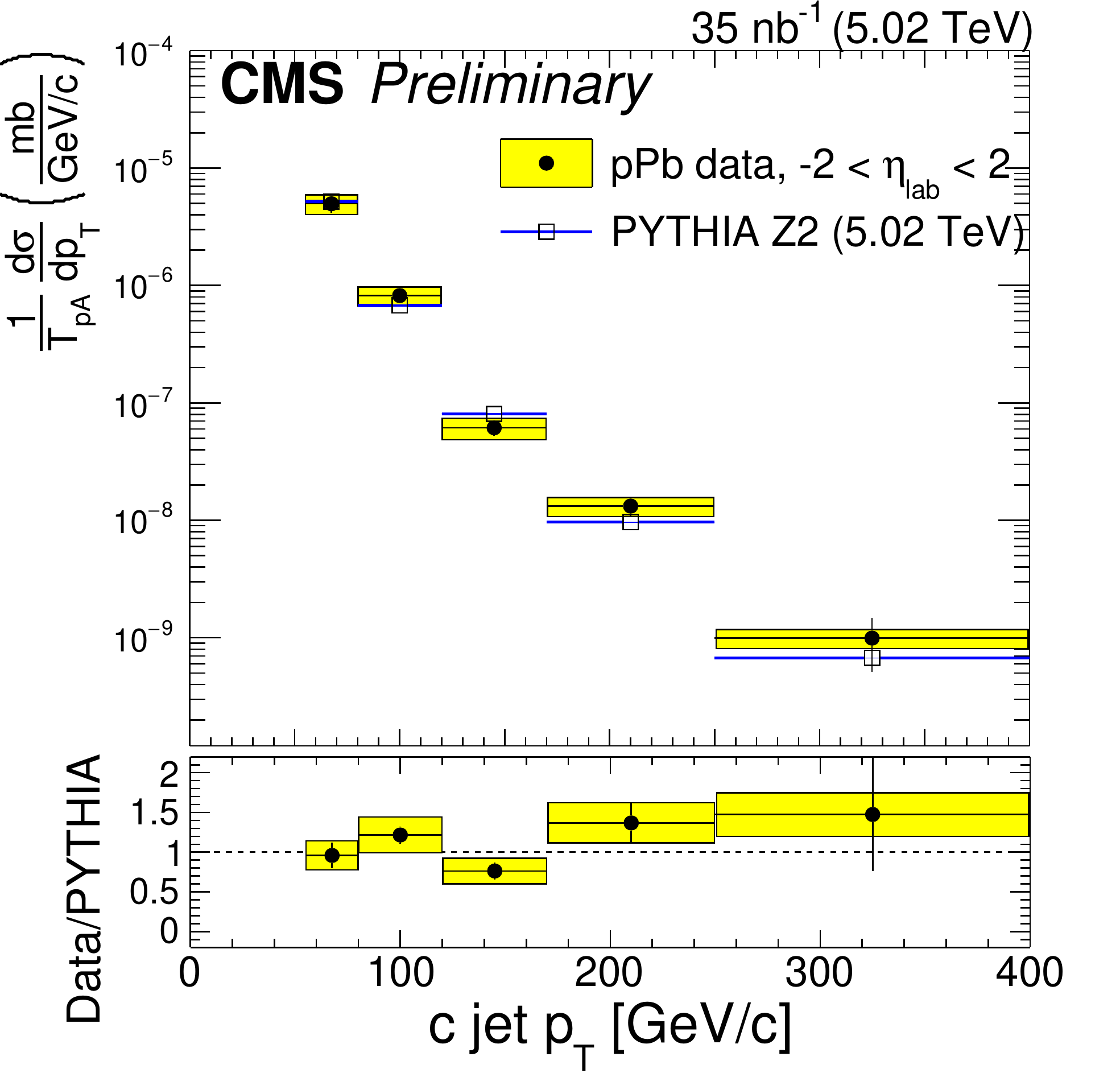}
\end{center}
\end{minipage}
\caption{(Left) b-jet \raa as a function of \pT within $|\eta|$ $<$ 2 in 0-10\% \PbPb collisions at \sqrtsnn = 2.76 TeV. The measurement of inclusive jet \raa is also shown for comparison. (Middle) b-jet $R^{PYTHIA}_{pA}$ as a function of \pT in 5.02 TeV \pPb collisions. (Right) c-jet invariant yield scaled by $T_{pA}$ as a function of \pT 5.02 TeV \pPb collisions. PYTHIA calculation is shown for comparison.}
\label{fig:hf_jet}
\end{figure}

Figure~\ref{fig:hf_jet} shows the CMS heavy flavor jets measurements. The b-jet reconstructed in $|\eta| <$ 2.0 in \PbPb collision is more suppressed in central collisions~\cite{bjet_PbPb}. In the left panel of the figure the \raa of b-jet and inclusive jet are compared and found to be consistent within uncertainties for \pT $>$ 80 \gevc. The middle panel shows the $R_{pA}$ measurements for b-jet as a function of \pT in 5.02 TeV \pPb collisions~\cite{bjet_pPb}. The result indicates that the b jet suppression measured in \PbPb collisions, is not due to CNM effects. In the right panel of the figure, the invariant yield of the first charm-jet measurement normalized by $T_{pA}$ as a function of \pT is compared with PYTHIA ~\cite{HIN-15-012}. As in the case of b-jet production, the data/PYTHIA ratio indicates no significant CNM effect in charm-jet production. 

\section{Summary and Outlook}
\label{summary_outlook}

The suppression of the three $\Upsilon$ states is found to be ordered, i.e. \raa ($\Upsilon (1S)$) $>$ \raa ($\Upsilon (2S)$) $>$ \raa ($\Upsilon (3S)$, for all centrality and \pT regions. This ordering is consistent with the expectation that a quarkonium state with weaker bounding energy is less likely to survive in the sQGP medium. However, the $\psi(2S)$ is found less suppressed than \jpsi in 3 $<$ \pT $<$ 30 \gevc, which may be caused by the coalescence production at different time scale during the medium evolution. The dependence of $\Upsilon$ production on charged particle multiplicity indicates final state effects in \pPb collisions. 

The prompt \DO \raa is consistent with charged particle \raa in 5.02 TeV \PbPb collisions, with the trend that \raa decreases with \pT for \pT $<$ 10 \gevc but increases toward higher \pT, which can be qualitatively described by many models. There are some indications that non-prompt\jpsi is less suppressed than prompt \DO  in intermediate \pT region, while at high \pT the b-jet \raa is consistent with the inclusive jet \raa within uncertainties. The $R_{pPb}$ measurements suggest that CNM effects do not have a large role in open heavy flavor \raa measurements in \PbPb. However, $R_{FB}$ measurements show significant CNM effects in non-prompt \jpsi production in \pPb collisions. 

At the time of writing this paper, many  new physics results in CMS are expected to be publicly available soon, part of which include \raa of fully reconstructed $B$ meson and non-prompt \DO meson, di-b-jet correlation, prompt \DO $v_N$. These new measurements should provide further inputs in understanding the heavy quark and sQGP medium interaction.


\section*{References}
\begin{thebibliography}{9}

\bibitem{matsui} T. Matsui, and H. Satz, Phys. Lett. {\bf B 178}, 416 (1986).

\bibitem{recomb} P. Braun-Munzinger and J. Stachel, Phys. Lett. {\bf B 490}, 196 (2000); R.L. Thews {\it et al.}, Phys. Rev. {\bf C 63}, 054905 (2001).

\bibitem{rad_eloss} M. Djordjevic {\it et al.}, Phys. Lett. {\bf B 632}, 81 (2006).

\bibitem{coll_eloss} H Van Hees {\it et al.}, Phys.  Rev. Lett. {\bf 100}, 192301 (2008); P.B. Gossiaux {\it et al.}, Nucl. Phys. {\bf A 830}, 203 (2009); X. Zhao {\it et al.}, Phys. Rev. {\bf C 82}, 064905 (2010); A. Buzzatti {\it et al.}, arXiv:1207.6020. 

\bibitem{coll_dissoc} R. Sharma {\it et al.}, Phys. Rev. {\bf C 80}, 054902 (2009).

\bibitem{ads_cft} W.A. Horowitz {\it et al.}, J. Phys. {\bf G 35}, 104152 (2008).

\bibitem{cgc} K. Tuchin, Nucl. Phys. {\bf A 904-905}, 627c-630c (2013).

\bibitem{HIN-12-014} CMS Collaboration, CMS PAS HIN-12-014 (2012).

\bibitem{ALICE_jpsi} B. Abelev {\it et al.}, Phys. Lett. {\bf B 734}, 314 (2014).

\bibitem{RHIC_jpsi} A. Adare {\it et al.}, Phys. Rev. {\bf C 84}, 054912 (2011); L. Adamczyk {\it et al.}, Phys. Lett. {\bf B 722}, 55 (2013).

\bibitem{seq_reg} X. Du, R. Rapp, Nucl. Phys. {\bf A 943}, 147 (2015).

\bibitem{UPC} CMS collaboration, arXiv:1605.06966 (2016). 

\bibitem{HIN-15-001} CMS collaboration, CMS PAS HIN-15-001 (2015).

\bibitem{JHEP04_103} CMS collaboration, JHEP {\bf 04}, 103 (2014). 

\bibitem{phnx_jpsi_dAu} A. Adare {\it et al.}, Phys. Rev. Lett. {\bf 107}, 142301 (2011).

\bibitem{Upsilon_polarization} CMS collaboration, arXiv:1603.02913 (2016).

\bibitem{HIN-16-001} CMS collaboration, CMS PAS HIN-16-001 (2016).

\bibitem{D0_models} J. Xu {\it et al.}, JHEP {\bf 02}, 169 (2016); M. Djordjevic {\it et al.}, Phys. Rev. {\bf C 92}, 024918 (2015); Shanshan Cao {\it et al.}, arXiv:1605.06447v1; T. Song {\it et al.},  Phys. Rev. {\bf C 93}, 034906 (2016); Z.-B. Kang {\it et al.}, Phys. Rev. Lett. {\bf 114}, 092002 (2015); Y.-T. Chien {\it et al.}, Phys. Rev. {\bf D 93}, 074030 (2016).

\bibitem{HIN-15-005} CMS collaboration, CMS PAS HIN-15-005 (2015).


\bibitem{B_RpPb} S. Chatrchyan {\it et al.} (CMS collaboration), Phys. Rev. Lett. {\bf 116}, 032301 (2016).

\bibitem{bjet_PbPb} S. Chatrchyan {\it et al.} (CMS collaboration), Phys. Rev. Lett. {\bf 113}, 132301 (2014).

\bibitem{bjet_pPb} S. Chatrchyan {\it et al.} (CMS collaboration), Phys. Lett. {\bf B 754}, 59 (2016).

\bibitem{HIN-15-012} CMS collaboration, CMS PAS HIN-15-012 (2015).


\end{thebibliography}
\end{document}